\documentclass[review]{elsarticle}

\usepackage{hyperref}
\usepackage{wasysym}
\usepackage{textcomp}
\usepackage{nicefrac}
\usepackage{upgreek}
\usepackage{booktabs}
\usepackage{verbatim}
\usepackage{multirow}

\usepackage{lineno,hyperref}
\modulolinenumbers[5]

\journal{Nuclear Instruments and Methods A}

\newcommand{\cm}{$\rm{cm}$}
\newcommand{\cmsq}{$\rm{cm}^2$}
\newcommand{\mm}{$\rm{mm}$}

\newcommand{\mum}{$\upmu \rm{m}$}

\begin{document}

\begin{frontmatter}

\title{High-Precision Contactless Optical 3D-Metrology \\ of Silicon Sensors}

\author[tue]{E. Lavrik\corref{cor1}}
\ead{Evgeny.Lavrik@uni-tuebingen.de}
\author[tue]{I. Panasenko}
\author[tue]{S. Mehta}
\author[gsi]{U. Frankenfeld}
\author[tue,gsi]{H.R. Schmidt}

\address[tue]{University of T\"ubingen, Auf der Morgenstelle 14, 72076 T\"ubingen, Germany}
\address[gsi]{GSI Helmholtzzentrum f\"ur Schwerionenforschung GmbH, Planckstrasse 1, 64291 Darmstadt, Germany}
\cortext[cor1]{Corresponding author}

\begin{abstract}

We describe a setup and procedures for contactless optical 3D-metrology of silicon micro-strip sensors. Space points are obtained by video microscopy and a high precision XY-table. The XY-dimensions are obtained from the movement of the table and pattern recognition, while the Z-dimension results from a Fast Fourier Transformation analyses of microscopic images taken at various distances of the optical system from the object under investigation. The setup is employed to measure the position of silicon sensors  mounted onto a carbon fibre structure with a precision of a few microns.
\end{abstract}


\end{frontmatter}

%\linenumbers

\section{Introduction}

Metrology, i.e., the precise knowledge of the position of tracking detectors,  is the key for a high quality data in experimental physics. We have, in the framework of the \textgravedbl Compressed Baryonic Matter\textgravedbl\ Experiment (CBM)~\cite{CBMBook2011}, developed advanced methods to determine the position of silicon sensors of a silicon tracker in all three dimensions relative to a reference point.
This knowledge constrains the parameter range significantly for later track-based alignment procedures, e.g., with programs like Millepede ~\cite{Milli}.
It also helps to develop and optimize the assembly procedure which should allow to position the modules within  $\approx$100~\mum\ of the nominal position.
It is as well possible to determine the warp of a sensor. This information has turned out to be useful when inspecting the electrical properties of a sensor on a probe station, because it allows to adjust the needle height (and thus the needle pressure onto the pad) according to the determined warp ~\cite{Panasenko2016, Panasenko2018}. It should be noted that the method described below requires a structured surface. This could be the strips and pads of a silicon micro-strip sensor or, e.g., the irregular structure of the micro-scratches of a polished aluminum surface. 

CBM is one of the four large experiments under construction at the future international accelerator center Facility for Anti-Proton and Ion Research (FAIR)~\cite{FAIRBTR, FAIRGP} in Darmstadt. The key detector of the CBM experiment is a complex, multi-layer Silicon Tracking System (STS)~\cite{STSTDR}. The performance requirements are, among others, a momentum resolution of better than 2\% at $p_t>$ 2$~GeV/c$ in a 1~Tm dipole magnetic field, and capabilities for the identification of particle decays with displaced vertices, e.g., those with strangeness content. To meet these requirements, it is necessary to know the position of the sensors with a precision equal or below the intrinsic resolution, which is determined by the strip pitch of 58~\mum.

The basic building block of the STS is a module, which consists of a double-sided silicon micro-strip sensor and an ultra-light multi-layer micro-cable which connects the sensor to the front-end electronics (FEE). The sensor sizes used in the setup are $6.2 \times 2.2$~\cmsq, $6.2 \times 4.2$~\cmsq, $6.2 \times 6.2$~\cmsq and $6.2 \times 12.4$~\cmsq. The micro-cables have a length of up to 50 cm, which is needed to mount the FEE outside of the detector's acceptance. Altogether, about 900 sensors are arranged to form eight tracking stations.

In this paper we briefly describe in Sect.~\ref{sec:setup} the mechanical and optical setup. Sect.~\ref{sec:procedures} contains a detailed description of the procedures applied to sets of sensor images to obtain the coordinates space points on the sensor surface relative to a reference point. In Sect.~\ref{sec:applications} we demonstrate the applicability of the described methods by the precision metrology of one the first assembled ladders of the CBM STS. Sect.~\ref{sec:summary} is a summary.

\section{Mechanical and Optical Setup}
\label{sec:setup}

The mechanical and optical setup which was employed to develop the software and procedures is described in detail elsewhere~\cite{Lavrik2017, Lavrik:2018osw}. The main components are  XY-linear motor stages using closed-loop feedback correction and a motorized zoom and focus assembly for the Z-stage. The optical components include a  12$\times$ zoom (0.58$\times$ - 7$\times$), a motorized 3~mm fine focus tube from Navitar\textsuperscript{\textregistered} and a 5 megapixel microscope camera from Motic\textsuperscript{\textregistered}.

\section{Position Measurements}
\label{sec:procedures}

After calibration, the XY-space points are simply extracted from motor positions and alignment marks, identified via pattern recognition, together with an appropriate calibration and conversion. This information can, for example, be used to determine the precise dimensions of a sensor (width, height), the distance of the sensor's edge to the alignment marks (needed if the alignment of the sensor is done relative to its edge), the parallelism of the strips to the sensor edge or, by turning the sensor upright, its thickness. The attainable precision depends on the XY-table accuracy and is in our case $\sim$ \mum\ (cf. \cite{Lavrik2017}). The determination of the Z-dimension (height) is based on a Fast Fourier Transform  (FFT) analysis of microscopic images of the sensor and is described below in more detail. 

\subsection{Principle of Height Measurement}

The precise positioning of the camera in Z-direction or change of the motorized focus stage value allows a height measurement of the object under inspection. This is done by analyzing the image sharpness at different motor or rather focus positions.

\begin{figure}[!htb]
	\centering
	\includegraphics[width=.8\columnwidth]{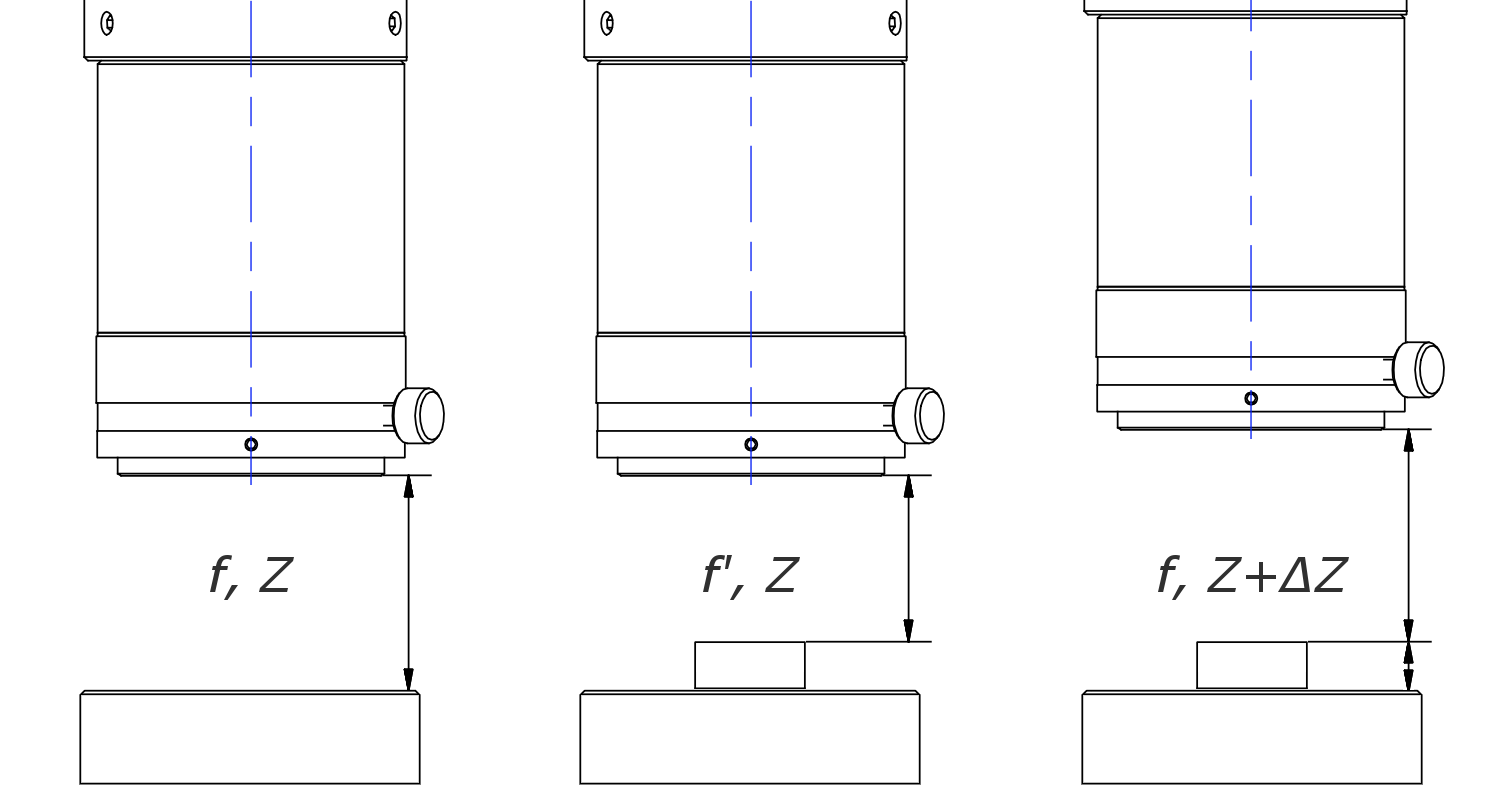}
	\caption{The object height measurement principle by adjusting the focus of the system in a calibrated fashion (middle panel) or by adjusting the vertical position of the optical system (right panel). Tube drawing taken from Navitar\textsuperscript{\textregistered}~\cite{Navitar}.}
	\label{fig:autofocus-principle}
\end{figure}

Fig.~\ref{fig:autofocus-principle} shows the principle of the height measurements. When inspecting a certain object, the focus of the system is first adjusted such that the image of the underlying surface is sharp and well focused (left panel of the figure). Then an object under test is put on the underlying surface, and the optical system is adjusted once again for the most focused image. This is done by adjusting the focus of the system (middle panel) or by adjusting the distance from the focal plane to the object by moving the whole system in an according Z-direction ($\Delta Z$ on the right panel). The calibrated change of the system focus is converted from motor to laboratory coordinates, which allows to extract the height of the object. We use motorized fine focusing (Fig.~\ref{fig:autofocus-principle}, middle) due to its higher precision compared to the movement of the whole, relatively heavy,  assembly (Fig.~\ref{fig:autofocus-principle}, right).

\subsection{Calibration of the Focus stage}
\label{sec:focus-calibration}

For the height measurements a precise calibration of the focus stage is mandatory. This was been done with a certified micrometer precision gauge block set from Mitutoyo Corp.~\cite{Mitutoyo}. The dependence of focus value vs. object height (from different combinations of the gauge blocks) is fit with a linear function yielding a slope coefficient k = 2.2333(3) motor steps/\mum, i.e., the motor steps to micrometer conversion ratio for the focus motor stage used. 

\begin{comment}
\begin{figure}[!htb]
	\centering
	\includegraphics[width=.7\textwidth]{pictures/Axis_Focus_Calibration.pdf}
	\caption{The autofocus values measured for different object heights.}
	\label{fig:axis-focus-calibration}
\end{figure}

%should not be called auto-focus value, better just: focus value

Fig.~\ref{fig:axis-focus-calibration} shows the measured focus value (in motor steps) for different gauge stack heights. 
\end{comment}

\subsection{Autofocusing}

The algorithmic implementation of an autofocusing procedure is done as follows: one steps over the range of motor positions and acquires the corresponding images with the camera. Every image is transformed with a FFT to the frequency domain\footnote{In the frequency domain every pixel represents the particular frequency contained in spatial domain.}. Then, as a measure for image sharpness, the sum of all frequencies of the complex image is calculated and associated with a particular motor position. The corresponding values are fit with a Cauchy-Lorentz distribution~\cite{Johnson1994} probability density function (cf. equation \ref{eq:cauchy-lorentz-distribution}):

\begin{equation}
\label{eq:cauchy-lorentz-distribution}
f(z; z_0,\alpha, \beta,\gamma) = \alpha + \beta \cdot \frac{1}{\pi\gamma \left[1 + \left(\frac{z - z_0}{\gamma}\right)^2\right]},
\end{equation}

\begin{figure}[!htb]
	\centering
	\includegraphics[width=.55\columnwidth]{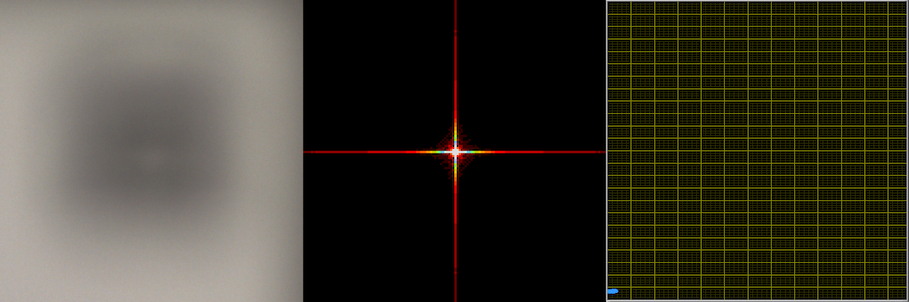}
	\rule[-0.3ex]{0pt}{1ex}
	\includegraphics[width=.55\columnwidth]{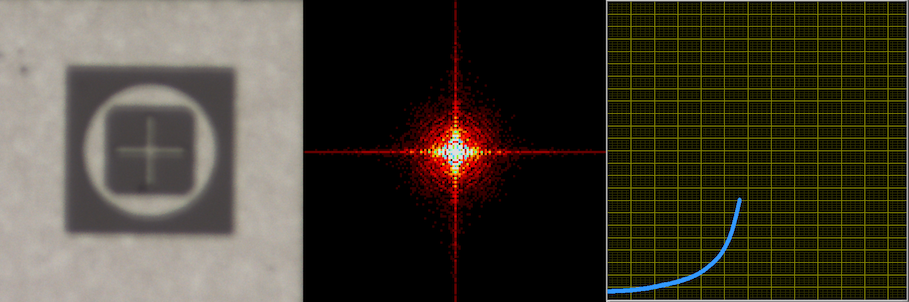}
	\rule[-0.3ex]{0pt}{1ex}
	\includegraphics[width=.55\columnwidth]{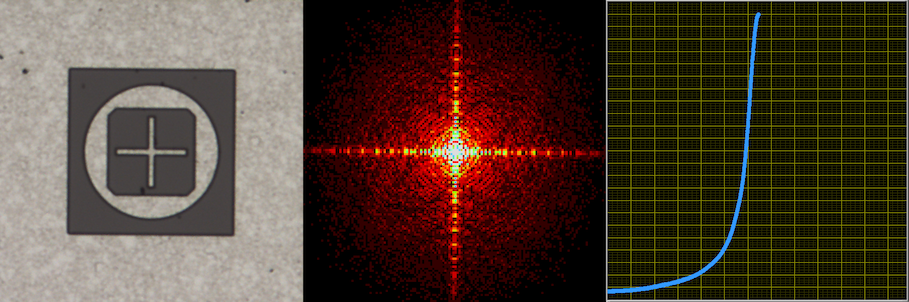}
	\rule[-0.3ex]{0pt}{1ex}
	\caption{An instant cut of the autofocusing process at 3 different motor positions, corresponding to non-focused image (top row), more focused image (middle row) and fully focused image (bottom row).}
	\label{fig:autofocusing-principle-steps}
\end{figure}

The variables $\alpha$, $\beta$, $\gamma$ and $z_0$ are the parameters to fit the distribution with the Levenberg-Marquardt~\cite{Lev01,Mar01} algorithm during offline analysis. The fit value $z_0$ corresponds then (within the extracted fit error interval) to the motor position of the most sharp image.

\begin{comment}
\begin{figure}[!htb]
	\centering
	\includegraphics[width=.7\columnwidth]{pictures/Autofocusing_Fitting.pdf}
	\caption{Image frequency response vs. motor step.}
	\label{fig:autofocus-fitting}
\end{figure}

Fig.~\ref{fig:autofocus-fitting} shows the distribution which has the most focused value around 3209 motor steps (roughly the mid interval of focus motor space) produced after scanning procedure with a step size of 10 motor steps.
\end{comment}

The procedure is visualized in Fig.~\ref{fig:autofocusing-principle-steps}. The left panel shows the image taken from the camera shows the sensor's alignment mark, the middle panel shows a graphical representation of a Fourier transformed image and the right panel shows the total amplitude distribution obtained from processing transformed images corresponding to the Lorentz distribution discussed above.

\subsection{Measurement Precision}

\begin{figure}[!htb]
	\centering
	\includegraphics[width=.7\columnwidth]{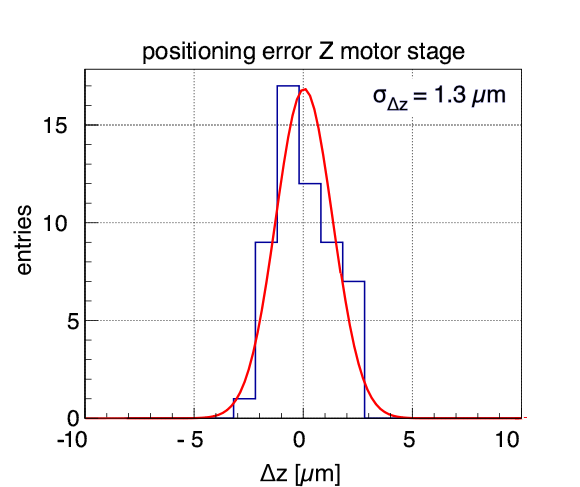}
	\caption{Estimation of the measurement precision from repetitive positioning.}
	\label{fig:autofocus-precision}
\end{figure}

To assess the precision obtained in measuring the height profile of an object evaluation, series of repetitive measurements were carried out. In these measurement the same height value was measured 50 times with a step size of 1 motor step to estimate the mean measured autofocusing value and its standard deviation. The result is shown in Fig.~\ref{fig:autofocus-precision}.

The tests were then repeated for different motor step sizes in the range from 1 to 500, i.e., 0.45~\mum\ to 225~\mum.

\begin{figure}[!hbt]
	\centering
	%\hspace*{\fill}
	\includegraphics[width=.45\textwidth]{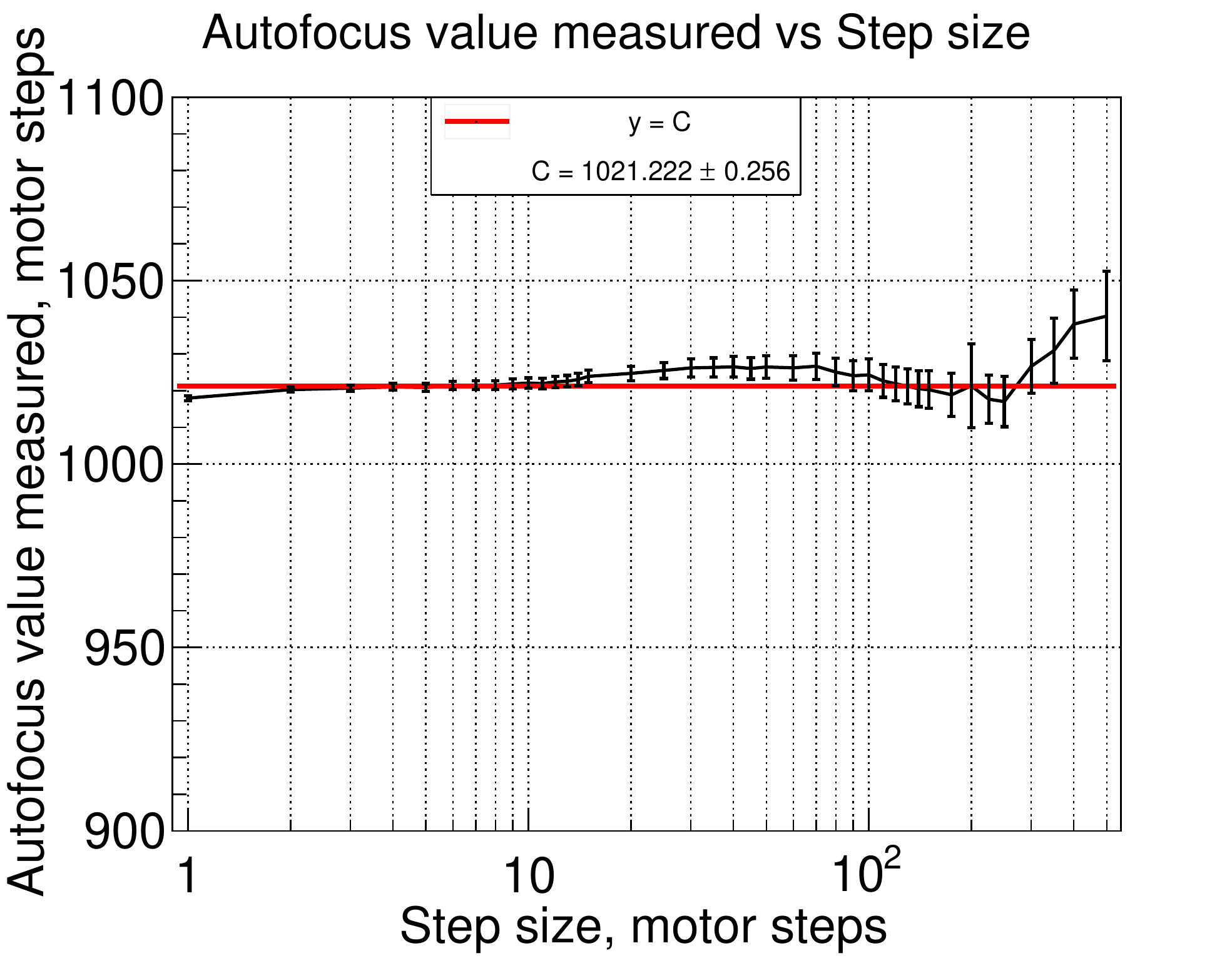}
	%\hfill
	%\includegraphics[width=.49\columnwidth]{pictures/Autofocusing_FitErr_vs_Step_I.pdf}
	%\hspace*{\fill}
	%\hspace*{\fill}
	\includegraphics[width=.45\textwidth]{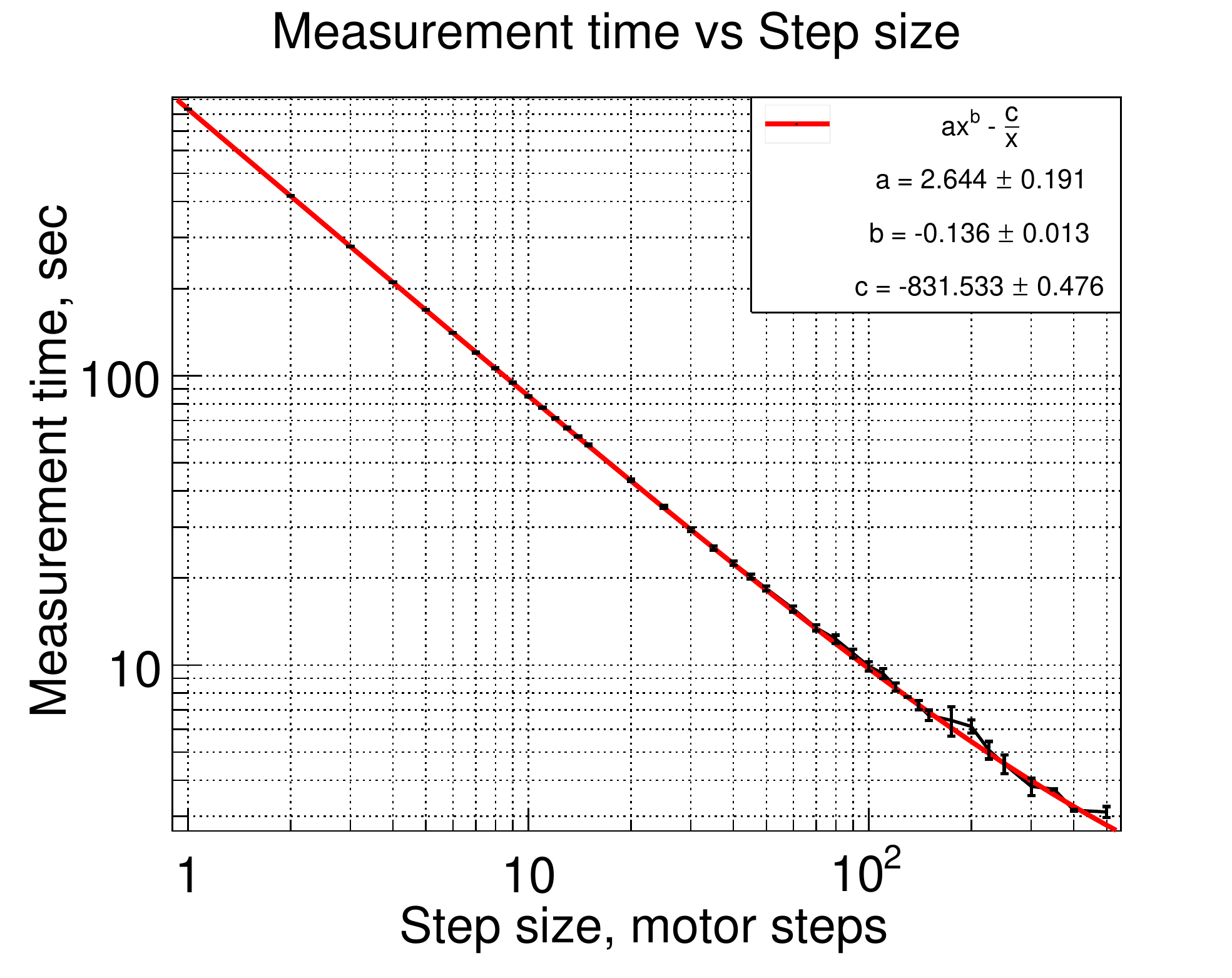}
	%\hspace*{\fill}
	\caption{Measured most focused value  (left), and measurement time (right) as function of the motor step size.}
	\label{fig:autofocus-val-step-I}
\end{figure}

Fig.~\ref{fig:autofocus-val-step-I} shows the autofocus value measured (left plot) and measurement time (right plot) as function of the motor stepping size. As can be seen a  smaller the motor step width gives, expectedly,  a more precise autofocus value. The precision, estimated in this way, is 1.3~\mum.

However, the time needed for a measurement increases exponentially with the stepping size. The time needed to scan the surface warp of a $6.2\times6.2$~\cmsq\ sensor with a precision of one motor step in Z-direction and a grid size of 1~\mm\ in the XY-plane would take about 174 hours. Thus, an optimization is needed to reduce the measurement time.

\subsubsection{Optimization}
\label{sec:warp-approach}

To speed up the time to find the most focused value, several methods of optimization are possible. The optimum method depends on the speed of the components of the system, e.g., zoom motor speed, camera FPS rate (frames taken per second) or CPU time needed to perform the FFT as well as on the specific measurement task. For scanning an extended area we have developed an adaptive approach which speeds up the process and as well increases the measurement precision, i.e., the step size, close to the maximum value of the distribution.
%It takes the value of the previously measured grid point into account .  The introduction of an \textit{approach parameter} allows to speed up the measurements: the motor step width is reduced by a certain factor, controlled by the approach parameter, as one comes close to the expected peak and increases by the same factor as the peak value is passed. 
This results in more statistics around the true peak, which improves the fitting. An additional introduction of a stop criterion that the current amplitude should not be less than $0.5$ of the maximum amplitude allows to exit the measurement loop at an early stage, thus saving more time. Details of the optimization procedure are found in ~\cite{Lavrik2017}.

At the highest precision measurement, i.e., with a motor step size of 1, the adaptive method is at least 8x faster for a single measurement. This reduces the total measurement time for the warp of a $6.2\times6.2$~\cmsq\ sensor to 21.3 hours for 1~mm steps in XY-directions. Any further decrease in measurement time will, however, reduce the measurement precision. In order to reduce the inspection time further other means, e.g. parallelization, have to be used.

\begin{comment}
					\paragraph{Method comparison}~\newline

Fig.~\ref{fig:autofocus-cmp} show the precision (in terms of fit errors) and single measurement time comparison.

\begin{figure}[!htb]
	\centering
	\hspace*{\fill}
	\includegraphics[width=.49\columnwidth]{pictures/Autofocusing_Comparison_FitErr.pdf}
	\hfill
	\includegraphics[width=.49\columnwidth]{pictures/Autofocusing_Comparison_Time.pdf}
	\hspace*{\fill}
	\caption{Method comparison for the stepping and adaptive approaches in terms of fit errors (left panel) and measurement times (right panel).}
	\label{fig:autofocus-cmp}
\end{figure}

Though being about 8 times faster for the adaptive method, the inspection time is still somewhat long. In order to optimize it and speed it up, one would need to give up the scanning granularity, since it is a quadratic dependence on the scanning step.

Fig.~\ref{fig:autofocus-full-insp-time} shows the total inspection time dependence from inspection granularity (the stepping size between measurements) for the approach parameter value of 2. The finer the measurement step, the better the inspection and more features could be seen on the surface. However, this means as well, that the inspection time increases dramatically.

\begin{figure}[!htb]
	\centering
	\includegraphics[width=.7\columnwidth]{pictures/Autofocusing_Full_Insp_Time.pdf}
	\caption{Log-log dependence of single measurement time vs. approach parameter value.}
	\label{fig:autofocus-full-insp-time}
\end{figure}

\end{comment}

\section{Applications}
\label{sec:applications}

As an example of power of the method we demonstrate the warp measurement of a  $6.2\times4.2$~\cmsq\ silicon sensor and, as a potential application, the metrology of sensors placed on CF ladders as they will be installed in the CBM experiment.
It should be noted the we employ pattern recognition algorithms to determine the  XY position of the sensor via alignment marks. The measurements can thus be automatized with minimal human intervention.

\subsection{Sensor Warp Measurement}

\begin{figure}[!htb]
	\hspace*{\fill}
	\includegraphics[width=.49\textwidth]{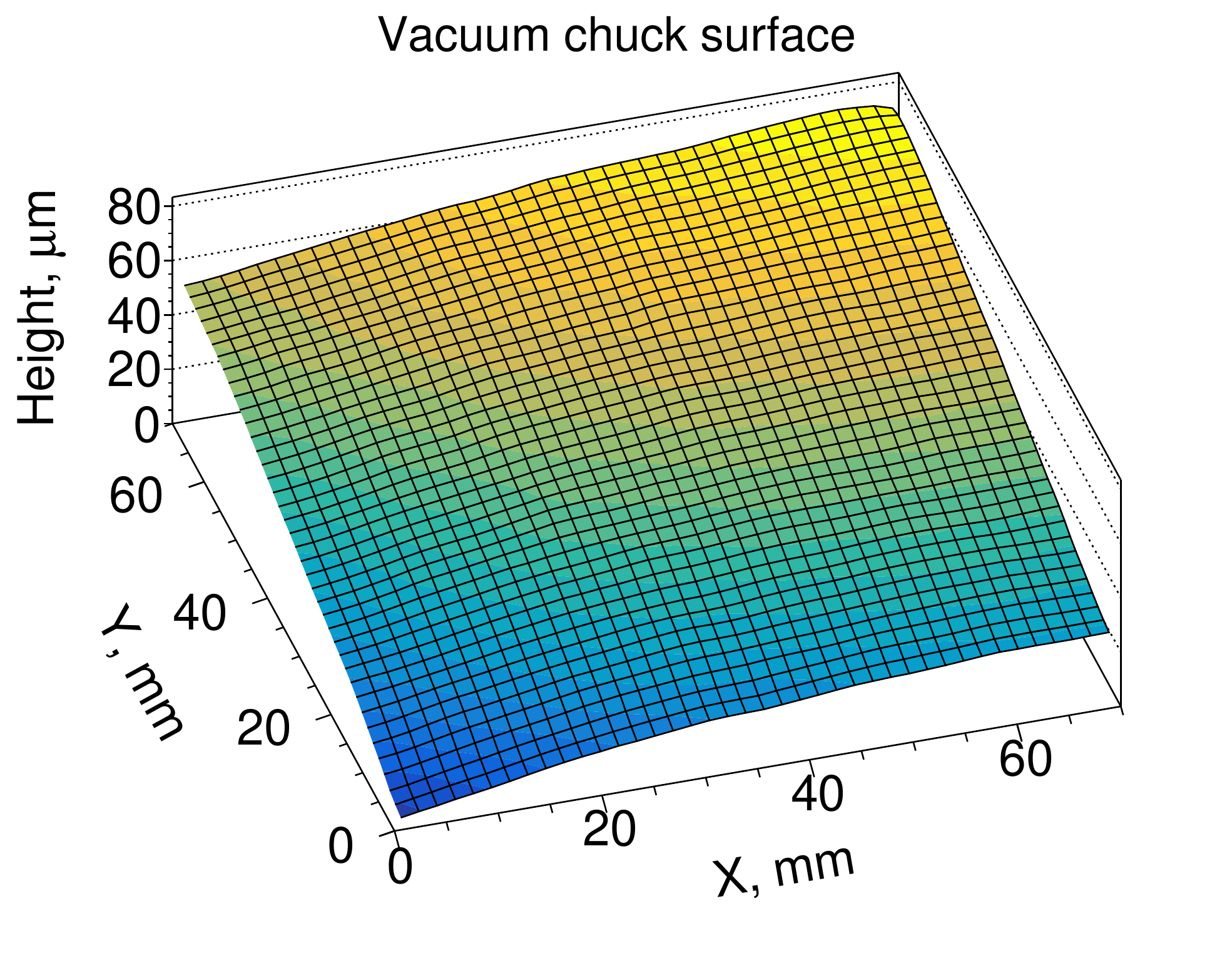}
	\hfill
	\includegraphics[width=.49\textwidth]{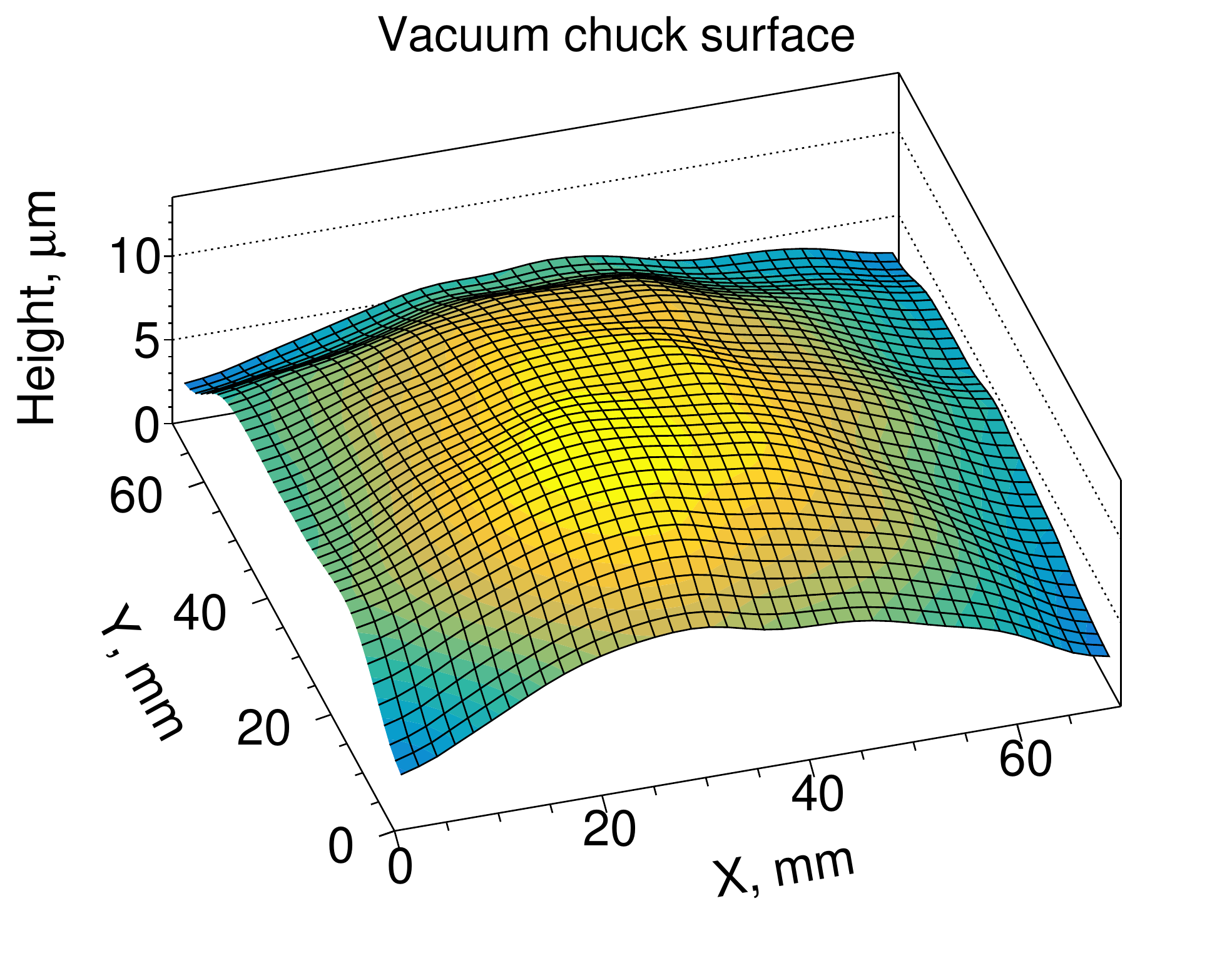}
	\hspace*{\fill}
	\caption{Height profile of a most central $7\times7$~\cm\ region of the vacuum chuck obtained with autofocusing method to measure object heights. The left panel shows uncorrected map. The right panel shows the corrected one.}
	\label{fig:chuck-surface}
\end{figure}

We evaluate corrections using the example of the warp measurement of silicon sensors. To obtain the proper height map of a sensor (warp) it is necessary to correct for any non-parallel surface of the sensors mount structure (granite table, vacuum chuck,  XY-sliding carriages etc.) with respect to the direction of movement in X and Y. To do so, a self-calibrating baseline height-measurement without sensor is carried out, which typically yields and inclined surface (cf. Fig.~\ref{fig:chuck-surface} left). This surface is fit with a 2D-plane, which is then subtracted from all subsequent measurement. Subtracting the 2D-plane from the baseline measurement itself might yield a residual structure (cf. Fig.~\ref{fig:chuck-surface} right). This structure could, for example, results from the mechanical machining precision of the assembly. Dependent on the metrology task or the envisioned precision, it could be included into the baseline correction. 
 
It should further be noted that knowing the Z-profile of, e.g., the vacuum chuck, is also important for the optical inspection, since, properly taken into account, it improves the movement over the sensor as it avoids focus loss of the images taken (cf. \cite{Lavrik2017, Lavrik:2018osw}).

The warp or Z-profile measurement has been successfully used in the electrical QA employing a custom made probe station~\cite{Panasenko2016}. Here, the height of the probe needle relative to the pad has been, based of the local z-position, adjusted such that a reliable contact between needle and pad is assured, but, at the same time, force onto the pad is minimized to avoid scratches on the pads.

\begin{figure}[!htb]
	\centering
	\includegraphics[width=.9\textwidth]{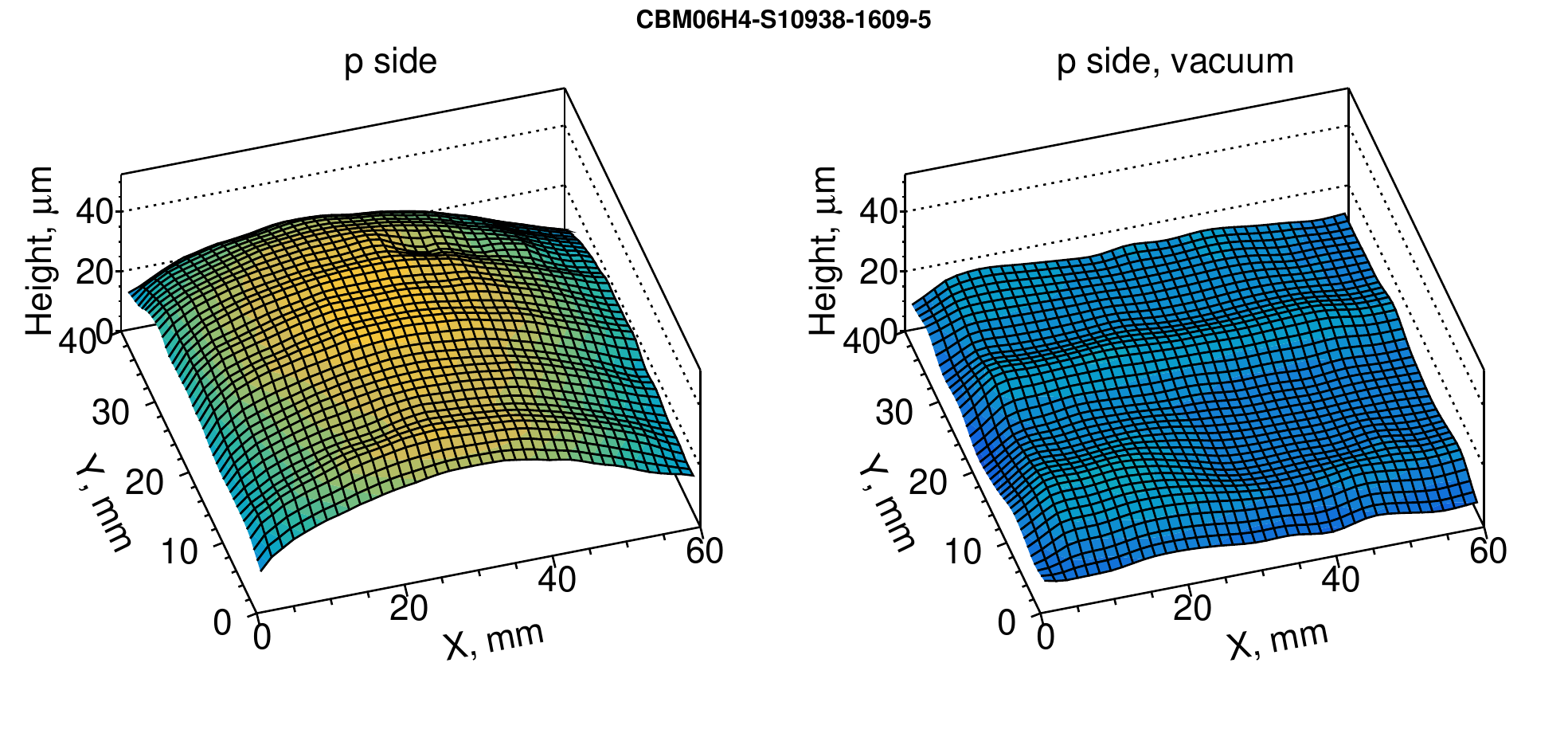}
	\caption{The warp of the prototype $6\times4$~\cmsq\ sensor measured without and with vacuum supplied to the chuck. The maximum warp value measured without vacuum is 44~\mum\ (left).  With vacuum applied residual structures of maximally 19~\mum\ are measured (right).}
	\label{img:warp-vac-effects}
\end{figure}

Finally, we show in Fig.~\ref{img:warp-vac-effects} the warp of a $6\times4$~\cmsq\ sensor measured without (left) and with vacuum (right) supplied to the chuck. The residual structure  in the left figure is (most likely) an artifact and demonstrates the limits of the method, which, however, is at the micron level. The residual structure  in the right figure is 
(most likely) the result of uneven suction of the vacuum chuck and demonstrates that, even of the vacuum table, the sensors are not completely flat and one recognizes the position of the vacuum holes of the chuck.

\subsection{Ladder Metrology}

\begin{figure}[!htb]
	\centering
	\includegraphics[width=.9\columnwidth]{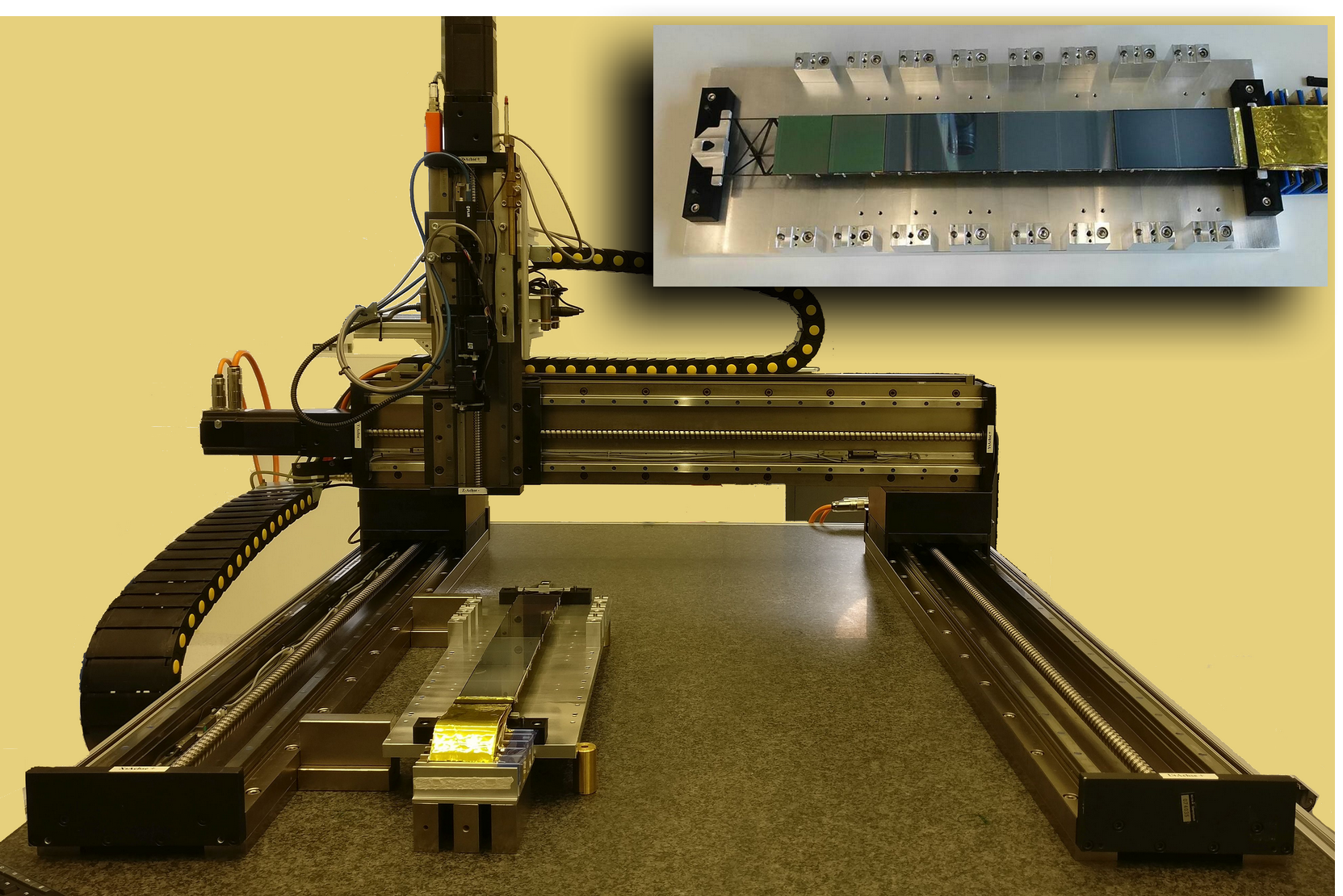}
	\caption{Setup to measure the position of sensors mounted onto long ladders. Shown are the large granite table, the X-, Y- and Z-motor stages and a prototype ladder with five sensors. The insert shows a top view of the ladder mounted in its assembly jig.}
	\label{fig:ladder metrology setup}
\end{figure}

Knowing the exact position of silicon sensors in the tracking detector is of paramount importance. The CBM experiment will use long ladders (up 100 cm) onto which up to 10 sensors will be glued with small "L-legs" \cite{Frankenfeld2017a}. To be able to measure the exact sensor position on the ladders after assembly, a large granite table and  X -Y motor bridges with traveling distances of 110  and 80 cm, respectively,  was acquired. The Z-stage is similar to the one described above. The whole assembly including a prototype ladder with five sensors mounted is shown in Fig.~\ref{fig:ladder metrology setup}. The overall precision, which includes, e.g., a correction of the granite table flatness and takes the long term reproducibility of the measurement into account, is $\approx 10~\mu$m.

The goal is to tune to assembly procedure such that the sensor positions do not deviate from the nominal position by more than 100~\mum.

\begin{figure}[!htb]
	\centering
	\includegraphics[width=.7\columnwidth]{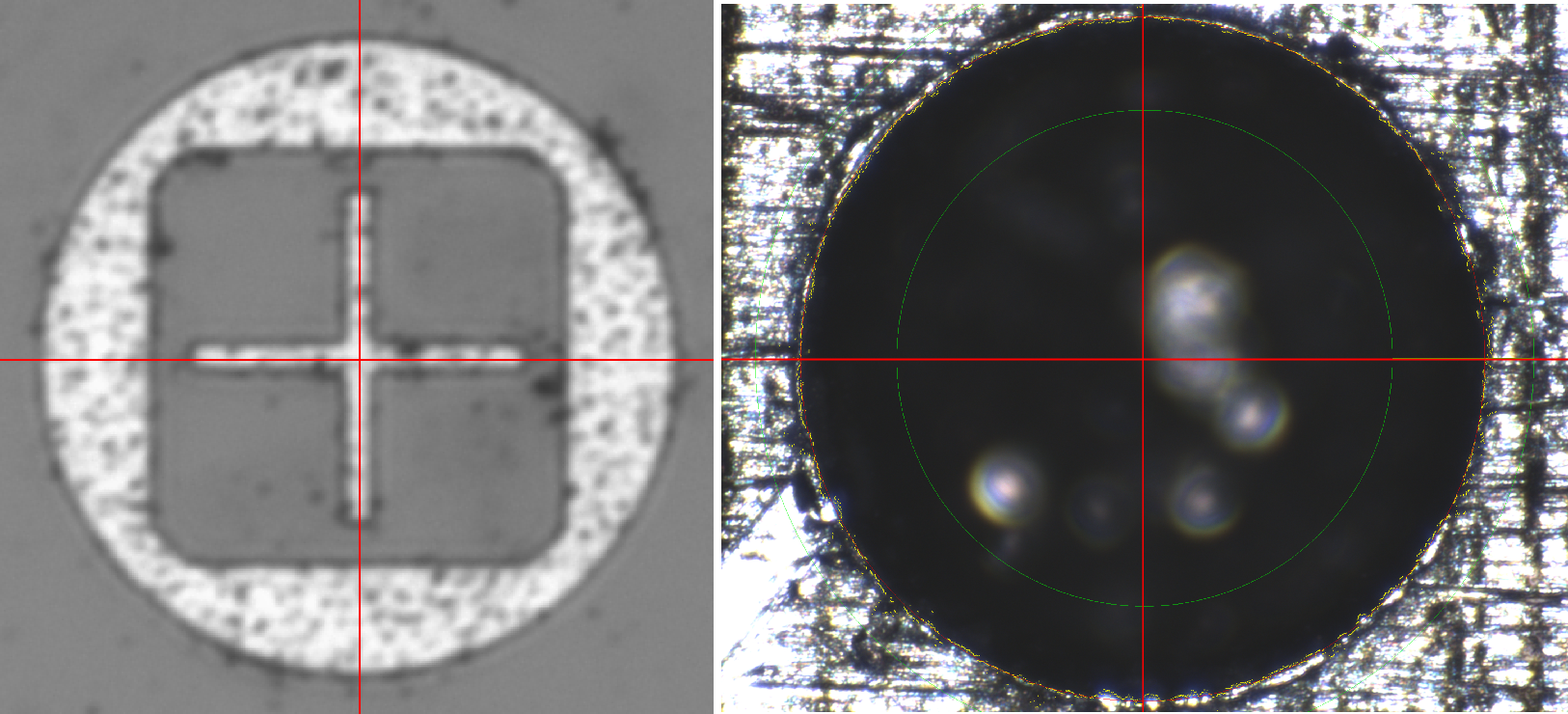}
	\caption{Left: Alignment mark on the sensor; right: blind home as reference mark on the assemble jig. The inner and outer ring give the search range for the edge. Note that the two images differ in scale by a factor of 10.}
	\label{fig:reference marks}
\end{figure}

The 3D-position of the sensors is determined from alignment marks on the sensor's surface in XY-plane, and from the Z-profile measurement described above (see Fig.~\ref{fig:reference marks}, left). The measurements of the space points on the sensors surface are done relative to reference marks on the assembly jig, which holds the ladder during assembly and during the position measurement. While the alignment marks on sensor have a sub-micron accuracy given by the precision of the sensor production masks, it is non-trivial task to make a reproducible "mechanical" mark on the assembly jig which can be recognized automatically by pattern recognition with the required micron precision. Our solution is a circular embossing of about 1~mm diameter. The edge of the circle is determined employing the  NI vision package. A circle is fit to many edge points which yields then the center with sufficient precision (see Fig.~\ref{fig:reference marks}, right).

\begin{figure}[!htb]
	\centering
	\includegraphics[width=1.0\columnwidth]{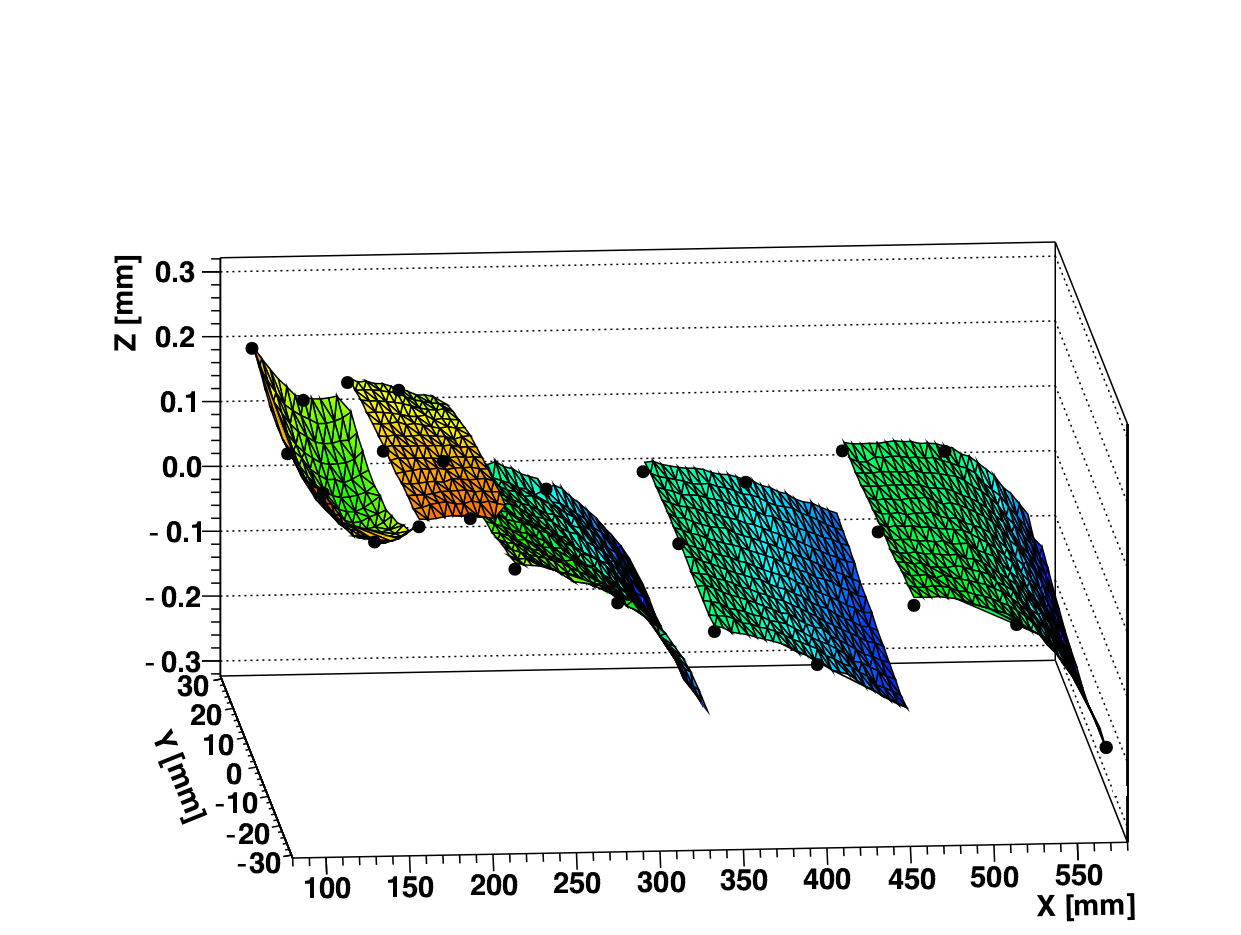}
	\caption{Measurement of the sensor surface space points with a step size of 5~mm. The black dots refer to the alignment marks on the sensor.}
	\label{fig:ladder space points}
\end{figure}

Having reference marks with sufficient precision, space points on the sensor's surface can be measured essentially with any grid size. The result of such a measurement is shown for one of the first assembled ladders in  Fig.~\ref{fig:ladder space points}. It should be noted that the staggering of the sensor position in Z by about 1~mm has been removed by software to allow sensible representation of the measured data. As can be seen, the sensor, glued to the CF ladder with L-legs, have still a sizable residual warp of the order of 10~\mum. On top the bending, the sensor surface have some inclination of the order 100-200~\mum, which reflects the current assembly precision in Z.  Table ~\ref{tab:scan result} shows the deviation in the lateral (X, Y) and rotational ($\Delta \phi$) degree of freedom from the nominal sensor position, where the nominal position is the one from the CAD drawing.

\begin{table}[!htb]
	\centering
	\begin{tabular}{c|c|c|c}
		 Sensor & $\Delta X [\mu m]$   & $\Delta Y [\mu m]$  	& $\Delta \phi [mrad]$                  \\ \hline
		1          & 287                     	& -45       				& -2.74           	\\
		2          & 151                      	& -68        				&  0.12            	\\
		3          & 40                  		& -42        				& 0.61             	\\
		4          & 56                   		& -107        				& -1.23               \\
		5          & 40                 		& -44     			 	& 0.96              	\\ 
	\end{tabular}
	\caption{Lateral and rotational deviation of the sensor from its nominal position.}
	\label{tab:scan result}
\end{table}

\begin{figure}[!htb]
	\centering
	\includegraphics[width=0.7\columnwidth]{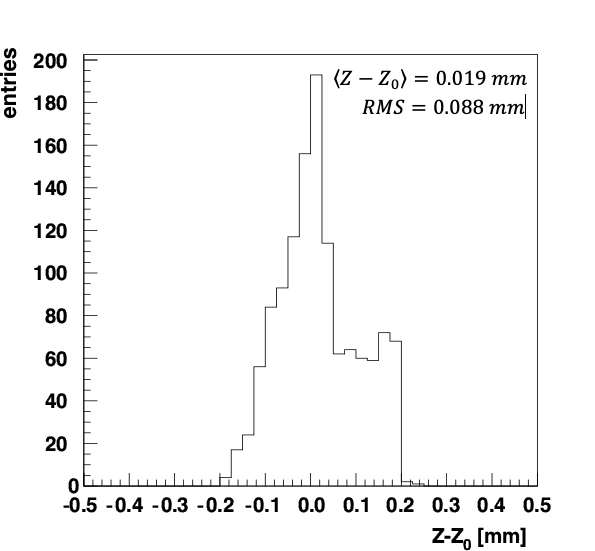}
	\caption{Histogram of the deviation of the measure Z-position from the nominal one.}
	\label{fig:ladder_space_pointsZ}
\end{figure}

The measured Z-positions are shown in Fig.~\ref{fig:ladder_space_pointsZ}. The RMS of the deviation from the nominal position is 88~\mum. 

As can be estimated from Table ~\ref{tab:scan result}, the measured positions in XY are, with 3 exceptions, already within the 100~\mum\ range which we think is acceptable. The accurate measurement of the 3D position will be used to further improve the assembly procedure.

%*********************

\section{Summary}
\label{sec:summary}

We have developed a setup and procedures which allows us to determine the position of an object in all 3 dimensions with very good ($\sim$~\mum) precision. The measurements are contactless. For the Z-direction (height) it is based on autofocusing via FFT, while for XY the motor positioning steps are used. We employ pattern recognition based on the NI vision package to recognize reference marks on the object under investigation, in our case silicon micro-strip sensors, which allows to carry the scans automatically. As an application we determined the position of silicon strip sensors glued onto a CF ladder. 
Here, the overall precision is about 10~\mum.
The method allows to check the precision of the mounting and gluing procedure of the sensors and to improve the procedure, if necessary. The measured sensor position on the ladder is further used as input for track-based software alignment.

\section*{Acknowledgements}
The work was performed under grant BMBF 05P16VTFC1. 

\section*{References}

\bibliography{literature/bibliography}

\end{document}